\documentclass[prl,twocolumn,showpacs,preprintnumbers,amsmath,amssymb]{revtex4}

\usepackage{dcolumn}
\usepackage{bm}

\begin{document}

\title{Unusual Properties of Superoscillating Particles}

\author{Achim Kempf}
\affiliation{
Department of Applied Mathematics, University of Waterloo\\
Waterloo, Ontario N2L 3G1, Canada}
\author{Paulo J. S. G. Ferreira}
\affiliation{
Dept. Electr{\'o}nica e Telecomunica{c}{\~o}es / IEETA\\
Universidade de Aveiro, 3810-193 Aveiro, Portugal}


\begin{abstract}
It has been found that functions can oscillate locally much faster than
their Fourier transform would suggest is possible - a phenomenon called
superoscillation. Here, we consider the case of superoscillating wave
functions in quantum mechanics. We find that they possess rather unusual
properties which raise measurement theoretic, thermodynamic and
information theoretic issues. We explicitly determine the wave functions
with the most pronounced superoscillations, together with their scaling
behavior. We also address the question how superoscillating wave functions
could be produced.
\end{abstract}

\pacs{03.65.Ta,03.67.-a,02.30.Nw}

\maketitle

\section{Introduction}
Consider a differentiable function which contains, according to its
Fourier transform, only wavelengths larger than some finite minimum
wavelength $\lambda_{\text{min}}$. Intuitively, one expects that this
function cannot oscillate on length scales significantly smaller than
$\lambda_{\text{min}}$. In fact, however, it has been found that such
functions are able to oscillate locally with wavelengths that are
arbitrarily smaller than $\lambda_{\text{min}}$. The phenomenon is known
as superoscillation, and indications of its occurrence have been discussed
in various contexts, from evanescent waves and seeming superluminal
propagation to the transplanckian problem of black holes, see
\cite{aha1}-\cite{ak-bh}. In the field of information theory, see e.g.
\cite{shannon,ferreira}, the phenomenon of superoscillating functions had
been observed even earlier, see \cite{levi}.

In this Letter, we investigate the rather unusual properties of quantum
mechanical wave functions which superoscillate. For example,
superoscillating particles tend to absorb energy from objects, thereby
acting as ``cooling-rays". Superoscillations come at a price, however,
which is that they are generally of small amplitude. Thus, our aim here is
to calculate and investigate the wave functions which possess the most
pronounced superoscillations, building on the methods of
\cite{ak-bh,levi,slepian}.

We find that the maximum amplitude of superoscillations decreases
polynomially with the frequency and exponentially with the duration of the
superoscillating stretch. We find that this scaling behavior is consistent
with information theoretic considerations. Finally, we propose a method by
which, in principle, superoscillating wave functions might be produced
experimentally.

To be precise, let us consider the set of wave functions with momentum
cutoff $p_\text{max}$:
\begin{equation} \psi(x) =
\frac{1}{\sqrt{2\pi\hbar}}\int_{-p_\text{max}}^{p_\text{max}}
\tilde{\psi}(p)~e^{\frac{ixp}{\hbar}}~dp\label{e1}
\end{equation}
Such wave functions are linear combinations of plane waves whose
wavelengths are at least as large as $\lambda_\text{min}=h/p_\text{max}$.
Nevertheless, as was shown in \cite{ak-bh}, such wave functions can
oscillate arbitrarily quickly on arbitrarily long stretches, in the
following sense: For any $N$ arbitrarily chosen points $\{x_k\}_{k=1}^N$
and arbitrarily chosen amplitudes $\{a_k\}_{k=1}^N$, there exist square
integrable wave functions $\psi$ which at the prescribed points take the
prescribed amplitudes
\begin{equation}
\psi(x_k)~=~a_k,~~~\text{for all} ~~~~ k=1,2,...,N\label{constraints}
\end{equation}
while also obeying the momentum cutoff expressed in Eq.\ref{e1}. Note that
$\psi$ will be differentiable, as is every wave function $\psi$ of the
form given in Eq.\ref{e1}. Note also that we are not yet normalizing
$\psi$.

If we choose the spacing of the points $\{x_k\}$ to be small compared to
the minimum wavelength, $x_{k+1}-x_k ~\ll~\lambda_\text{min}$, while
choosing oscillatory amplitudes $\{a_k\}$, e.g. $a_k=(-1)^k$, then we say
that the resulting wave function $\psi$ is superoscillating. Intuitively,
the superoscillations do not show as high frequencies in the Fourier
transform because of subtle cancellations in the Fourier integration over
all of the wave function, i.e. over both its superoscillating and
non-superoscillating parts. We will find that to this end the amplitudes
in the superoscillating part of the wave function must be significantly
smaller than the amplitudes in the non-superoscillating parts.
\section{Maximal superoscillations}
Our aim, therefore, is to calculate those normalized wave functions which
possess the most pronounced superoscillations. The strategy will be to use
a variational principle to first calculate that wave function $\psi$ of
the form of Eq.\ref{e1} which possesses the smallest norm while obeying
the arbitrarily chosen constraints of Eqs.\ref{constraints}. Second, we
normalize the wave function. In this way, having minimized the norm of
$\psi$, we insure that after the normalization of $\psi$ the amplitudes of
the superoscillations are maximal.

Explicitly, we minimize the norm $\int_{-\infty}^\infty
\psi^*(x)\psi(x)~dx = \int_{-p_\text{max}}^{p_\text{max}}
\tilde{\psi}^*(p)\tilde{\psi}(p)~dp$~ of the wave function $\psi$, subject
to the constraints of Eq.\ref{constraints}, which can also be written as:
\begin{equation}
\frac{1}{\sqrt{2\pi\hbar}}~
\int_{-p_\text{max}}^{p_\text{max}}\tilde{\psi}(p) ~ e^{\frac{i x_k
p}{\hbar}}~dp ~=~ a_k \label{conmom}
\end{equation}
Thus, with Lagrange multipliers $\mu_r$, the Lagrangian of the variational
problem reads
\begin{eqnarray*}
L & = & \int_{-p_\text{max}}^{p_\text{max}}
\tilde{\psi}^*(p)\tilde{\psi}(p)~dp\\
& & +~\sum_{r=1}^{N}~\frac{\mu_r}{\sqrt{2\pi\hbar}}
\int_{-p_\text{max}}^{p_\text{max}} \tilde{\psi}(p)~e^{\frac{i x_r
p}{\hbar}} ~dp ~~+~~ c.c.,
\end{eqnarray*}
from which we obtain:
\begin{equation}
\tilde{\psi}(p)
~+~\sum_{r=1}^N~\frac{\mu^*_r}{\sqrt{2\pi\hbar}}~e^{\frac{-i
x_rp}{\hbar}}~=~0 \label{moment}
\end{equation}
Using Eq.\ref{conmom}, we have therefore:
\begin{equation}
\frac{-1}{2\pi\hbar}~ \sum_{r=1}^N~\mu^*_r
\int_{-p_\text{max}}^{p_\text{max}}
~e^{\frac{ip}{\hbar}(x_k-x_r)}~dp~=~a_k\label{sol}
\end{equation}
We define the symmetric matrix $\{S_{k,r}\}_{k,r=1}^N$ through:
\begin{eqnarray}
S_{k,r} & = & \frac{1}{2\pi\hbar}~\int_{-p_\text{max}}^{p_\text{max}}~
e^{\frac{ip}{\hbar}(x_k-x_r)}~dp\nonumber\\ \\
 & = & \frac{\sin\left((x_k-x_r)
 p_\text{max}/\hbar\right)}{\pi(x_k-x_r)}\nonumber \label{prolatedef}
\end{eqnarray}
Note that $S$ is invertible since it is positive definite:
\begin{equation}
\sum_{r,k=1}^Nv^*_r S_{r,k}v_k = \frac{1}{2\pi\hbar}
\int_{-p_\text{max}}^{p_\text{max}}\left\vert\sum_{l=1}^N v_l~
e^{ipx_l}\right\vert^2 dp
\end{equation}
Eq.\ref{sol} now takes the form
\begin{equation} -\sum_{r=1}^N~S_{k,r}~\mu^*_r~=~a_k,
\end{equation}
and we can solve for the coefficients $\mu^*_r$:
\begin{equation}
\mu^*_r~=~-\sum_{m=1}^N~S^{-1}_{r,m}~a_m\label{ma}
\end{equation}
Using Eq.\ref{moment}, we finally obtain the desired wave function in the
momentum space representation:
\begin{equation}
\tilde{\psi}(p)~=~\frac{1}{\sqrt{2\pi\hbar}}~\sum_{r=1}^N~\sum_{m=1}^N~
S^{-1}_{r,m}~a_m ~e^{\frac{-i x_rp}{\hbar}}
\end{equation}
In the position representation, the superoscillating wave function reads:
\begin{eqnarray}
 \psi(x) & = & \frac{1}{2\pi\hbar}
~\int_{-p_\text{max}}^{p_\text{max}}~\sum_{r=1}^N~\sum_{m=1}^N~
S^{-1}_{r,m}~a_m ~e^{\frac{i(x -x_r)p}{\hbar}}~dp\nonumber\\\nonumber \\
  & = & \sum_{r,m=1}^N~S^{-1}_{r,m}~a_m~\frac{
  \sin\left((x-x_r)p_\text{max}/\hbar\right)}{\pi(x-x_r)}\label{relu}
\end{eqnarray}
 We see that $\psi$ is a delicate linear combination of shifted copies
of the function $\sin(2\pi x/\lambda_\text{min} )/x$ each of which is of
only slow variation. Among all wave functions which obey Eq.\ref{e1} and
which take the prescribed amplitudes $a_k$ at the $N$ prescribed positions
$x_k$, the wave function given in Eq.\ref{relu} possesses the smallest
norm. Thus, the normalized wave function $\psi^{(n)} = \frac{1}{\vert\vert
\psi\vert\vert}~\psi$ superoscillates with the largest achievable
amplitudes:
\begin{equation}
\psi^{(n)}(x_k)~=~\frac{a_k}{\vert\vert \psi\vert\vert},~~~~~\text{for
all~~} k=1,...,N \label{21}
\end{equation}
\section{Self-acceleration through single slit}
The properties of superoscillating wave functions are unusual:

Consider a low-momentum particle as it passes through a narrow slit in a
screen. The particle's position in the direction parallel to the screen
thereby becomes determined to within the width $L$ of the slit. By the
uncertainty principle, this implies that the particle's momentum $\Delta
p$ parallel to the screen becomes uncertain, such as to obey $\Delta p\ge
\frac{\hbar}{L}$. Thus, as is well-known, a particle may acquire momentum
when passing through a narrow slit.

We can now see, however, that the uncertainty principle is not the only
reason why particles can acquire momentum when passing through a slit.
Namely, consider an incident low-momentum particle whose wave function in
the direction parallel to the screen possess spatial superoscillations
just where the wave function hits the slit in the screen.

In this case, the wave function which emerges from the slit is spatially
oscillating with the very short wavelength of the superoscillations where
the slit is and is zero elsewhere. The emerging wave function's rapid
oscillations now do show as high frequencies in its Fourier transform.
This is because the emerging wave no longer possesses the high-amplitude
non-superoscillating parts which cancelled the high frequencies'
occurrence in the incident wave's Fourier transform. Thus, the emerging
quantum particle acquires a correspondingly high momentum expectation
value.

Crucially, since the superoscillating stretch of the particle's wave
function can be chosen arbitrarily wide, also the width of the slit can be
chosen arbitrarily wide. This shows that this new mechanism by which
particles are accelerated when passing through a slit occurs independently
from and on top of the uncertainty relation.
\section{Scaling behavior}
Superoscillations come at a price, however, which is that in normalized
wave functions any superoscillations are generally of very small
amplitude, implying that particles are not very likely to be found in
regions where their wave function is superoscillating.

To see this, note that due to Eq.\ref{21}, the superoscillations'
amplitude is suppressed by $\vert\vert\psi\vert\vert$, for which
Eq.\ref{moment} yields:
\begin{eqnarray*}
\vert\vert \psi\vert\vert^2 & = & \frac{1}{2\pi\hbar}
\int_{-p_\text{max}}^{p_\text{max}}~\left\vert\sum_{r=1}^N~\mu^*_r~
e^{\frac{-i x_r p}{\hbar}}\right\vert^2~dp\\
 & = & \sum_{k,r=1}^N ~\frac{\mu_r^*~\mu_k}{2\pi\hbar}
 ~\int_{-p_\text{max}}^{p_\text{max}}
 ~e^{\frac{i(x_k-x_r)p}{\hbar}}~dp\\
  & = & \vec{\mu}^\dagger ~S~\vec{\mu},
\end{eqnarray*}
We used vector notation: $\vec{\mu}=\{\mu_r\}_{r=1}^N$. Using Eq.\ref{ma},
we obtain:
\begin{equation}
\vert\vert \psi\vert\vert^2    ~=~ \vec{a}^\dagger
~S^{-1}~\vec{a}\label{norm}
\end{equation}
Clearly, the largest suppression occurs when $\vec{a}$ is an eigenvector
of the symmetric matrix $S^{-1}$ with the largest eigenvalue, i.e. when
$\vec{a}$ is an eigenvector of the symmetric matrix $S$ with the smallest
eigenvalue, say $s_\text{min}$. The normalized wave function $\psi^{(n)}$
then superoscillates with the amplitudes:
\begin{equation}
\psi^{(n)}(x_k)~=~ s^{1/2}_\text{min}~\frac{a_k}{\vert\vert
\vec{a}\vert\vert}
\end{equation}
Explicitly, let us prescribe the superoscillating amplitudes at
equidistant points with spacing $\Delta x$, namely $x_k = k\Delta x$ for
$k=0,...,(N-1)$. In this case, we obtain a matrix $S$ whose eigenvalues
were studied in \cite{slepian}: The definition of the prolate matrix
$\rho(N,W)$ in Eq.21 of \cite{slepian} matches ours through $S_{r,k}=
\rho_{r,k}(N,W(\Delta x))/\Delta x$ with $W(\Delta x)=\Delta x
p_\text{max}/h$.

Eq.64 of \cite{slepian} then straightforwardly yields the scaling of
$s_\text{min}$, and thus of $\psi^{(n)}(x_k)$, for fixed $N$ and
decreasing spacing $\Delta x$:
\begin{equation}
\psi^{(n)}(x_k)\propto s^{1/2}_\text{min} \propto (\Delta
x)^{N-1}\label{poly}
\end{equation}
Thus, if a fixed number $N/2$ of equidistant superoscillations is
compressed into a smaller region, i.e. if the superoscillations'
wavelength $2\Delta x$ is reduced, then the superoscillation amplitude
decreases polynomially with the wavelength of the superoscillations.

In the case where the superoscillation wavelength, i.e. the spacing
$\Delta x$, is held fixed and the number $N$ of superoscillations is
increased, we can use Eqs.13,58 from \cite{slepian} to readily find the
scaling behavior of the smallest eigenvalue $s_\text{min}$ of $S$ for
large $N$, to obtain:
\begin{equation}
\psi^{(n)}(x_k)\propto s^{1/2}_\text{min} \propto N^{1/4} e^{-\gamma N/2}
\label{exp}
\end{equation}
Here, $\gamma$ is positive and depends on $\Delta x$ but not on $N$. Thus,
we proved that the amplitude of superoscillations decreases exponentially
with the number of superoscillations, i.e. with the length of the
superoscillating stretch, as had in fact been conjectured by Berry, in
\cite{berry1}.

While these scaling results imply exceedingly small superoscillation
amplitudes, we note that if the superoscillating stretch of a wave
function does pass through a slit, then the emerging superoscillations'
amplitudes get boosted to normal levels through the renormalization of the
wave function.

\section{Open questions}
The phenomenon of superoscillations raises a number of fundamental
questions, which we here only begin to address:

1) In the single slit setup of above, the superoscillating particles which
do pass through the slit will gain arbitrary predetermined amounts of
momentum and energy. Due to momentum and energy conservation, this gain
must be balanced by a corresponding loss in the momentum and energy of the
screen. Thus, those particles are accelerated by strong interactions with
the slit walls, which are in effect cooling the screen. Indeed, wave
functions can act as ``cooling rays" where they superoscillate. The energy
picked up by those particles is available to do work because the
particle's motion after the screen is directed, since their momentum is
largely predetermined. Consistency with the second law of thermodynamics
therefore requires either a) that the cooling effects of superoscillating
parts of the wave functions are necessarily offset by the presence of a
heating non-superoscillating part or b) that it is necessarily
entropy-expensive to produce superoscillating wave functions.

2) Further, the rate at which incident particles pass through the slit and
thereby take energy from the screen does not depend on the screen's
temperature. How, therefore is it ensured that the screen always has
sufficient energy and momentum available for those particles which happen
to pass through the slit? The answer should be, although we do not prove
this here, that if the position of a cold screen is known sufficiently
accurately to ensure that it is the superoscillating part of incident wave
functions which passes through the slit, then by the uncertainty relation
the screen possesses sufficient momentum uncertainty to be able to provide
the required momentum to the superoscillating particles which pass through
the slit. Conversely, if the screen is known to possesses very little
momentum, then by the uncertainty relation the position of the screen and
its slit cannot be known with sufficient precision to ensure for the
emerging particles that it was the superoscillating part of their wave
function which passed through the slit.

3) Generally, one may expect that in a single-particle experiment, if a
property of the particle can be predicted, the particle ought to have
possessed that property before the measurement. However, we now know for
example that if a low-momentum particle is found in a region where its
wave function is rapidly superoscillating then it will be found to possess
a very large and predictable momentum. Nevertheless, the particle clearly
did not possess this large momentum beforehand and only acquires it in the
measurement process.

4) Continuing this argument, one might expect that a superoscillating
stretch of a wave function could act as a false source of gravity. Namely,
a delocalized particle might locally gravitate and be curving space
according to the scale of energy-momentum density that it would possess if
the particle was found in that local region, rather locally gravitating
according to its actual energy-momentum if there is no measurement. Such
false sources of gravity should be suppressed effectively by the fact that
the amplitudes in superoscillating stretches of wave functions are very
low, as we saw.

5) Interestingly, our finding that the superoscillations' amplitude
decreases exponentially with the length of the superoscillating stretch
becomes plausible after translation into the language of information
theory. Namely, instead of wave functions with a finite momentum cutoff,
consider signals with a finite bandwidth. We see that by using
superoscillating signals it is in principle possible to encode arbitrary
amounts of information into a short interval of a low-bandwidth signal
which then passes through a low-bandwidth channel of some fixed finite
noise level. However, as Shannon showed in \cite{shannon}, signals which
pass through a channel of bandwidth $\omega_\text{max}$ can deliver
information at most at a rate $\omega_\text{max} \log_2(1+S/N)$, where
$S/N$ is the signal-to-noise ratio. Thus, the price for superoscillatory
information compression in signals is an exponential increase in the
required signal power, which is consistent with our finding of an
exponential increase of the functions' norm. This example suggests that if
there is an actual short-wavelength cutoff for physical fields in nature,
say at the Planck scale, it may usefully be thought of as a finite upper
bound on information density.

6) The question arises how superoscillating wave functions might be
produced experimentally. For example, can we design a potential $V(x)$
whose ground state has a superoscillating stretch? One possibility might
be to start with a harmonic oscillator potential centred about $x=0$. Its
ground state's momentum range is effectively cut off at high momenta since
it behaves as $\propto e^{-kp^2}$. Let us add to the potential several
relatively sharp spikes in close proximity, close to $x=0$. The ground
state's wave function should then become quickly varying in the region
where the spikes are, while also suppressed in amplitude. Whether this
would effectively constitute a superoscillatory ground state remains to be
studied.

\bf Acknowledgement: \rm A.K. and P.J.S.G.F. were in part supported by the
Natural Sciences and Engineering Research Council of Canada (NSERC) and by
the Funda\c{c}\~ao para a Ci\^encia e Tecnologia (FCT) of Portugal
respectively.

\end{document}